\documentclass[sigconf]{acmart}
\AtBeginDocument{%
  }

\begin{document}

\title{Stereoscopic Cylindrical Screen (SCS) Projection}

\author{Lim Ngian Xin Terry}
\affiliation{%
  \institution{Carnegie Mellon University}
   \country{USA}
}
\email{terrilim@cmu.edu}

\author{Adrian Xuan \\ Wei Lim}
\affiliation{%
  \institution{Pittsburgh}
   \country{USA}
}
\email{adrian.lim@u.nus.edu}

\author{Ezra Hill}
\affiliation{%
  \institution{Pittsburgh}
   \country{USA}
}
\email{ezra@mobiusteapot.com}

\author{Lynnette Hui \\ Xian Ng}
\affiliation{%
  \institution{Carnegie Mellon University}
   \country{USA}
}
\email{lynnetteng@cmu.edu}

\renewcommand{\shortauthors}{Terry et al.}

\begin{abstract}
We present a technique for Stereoscopic Cylindrical Screen (SCS) Projection of a world scene to a $360^0$ canvas for viewing with 3D glasses. To optimize the rendering pipeline, we render the scene to four cubemaps, before sampling relevant cubemaps onto the canvas. For an interactive user experience, we perform stereoscopic view rendering and off-axis projection to anchor the image to the viewer. This technique is being used to project virtual worlds at CMU ETC, and is a step in creating immersive viewing experiences.
\end{abstract}


\begin{teaserfigure}
  \includegraphics[width=\textwidth]{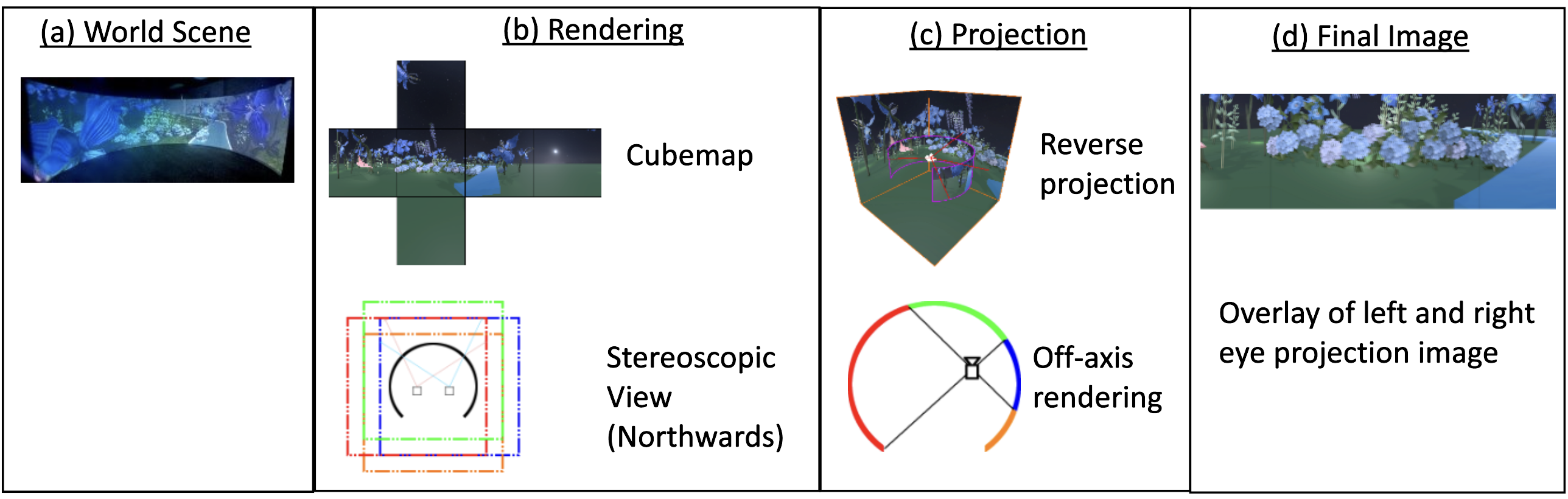}
  \caption{Stereoscopic Cylindrical Screen (SCS) Projection Pipeline}
  \label{fig:Framework}
\end{teaserfigure}


\maketitle

\section{Introduction}
Cave Automatic Virtual Environment (CAVE) systems have several advantages over HMD-based VR, primarily for multi-user use cases\cite{lebiedz2021multiuser}. CAVE systems render onto surrounding walls, enabling users to maintain visual awareness of one another within the shared physical environment. CAVEs using a cylindrical screen offer a wide field of view and more effective space for concurrent users than planar CAVEs. Many planar CAVEs render a stereoscopic 3D image to users, providing depth cues and helping with immersion for art and entertainment experiences\cite{lo2003stereo}. Planar caves create a parallax effect by rendering each image with an offset per eye. This rendering technique is not viable for cylindrical captures stereo depth separation will appear inverted near the center antipodal pair \cite{peleg2001omnistereo}.

Past work generated stereoscopic panoramic images for CAVEs and VR scenes. Many of these techniques are suitable for pre-rendered video, but not computationally viable for real-time or head-tracked rendering. This includes image-based geometric wrapping techniques \cite{karthik2024efficient}, equirectangular projection for $360^0$ views \cite{marrinan2021real}, object-warping and multiple-view methods for $240^0$ views \cite{simon2004omnistereo}.

Our Stereoscopic Cylindrical Screen (SCS) projection approximates accurate depth rendering to a wraparound view of a cylindrical screen. SCS allows a stereo effect to be visible by one or more users from any possible viewing angle. We optimize SCS projection through rendering a world scene onto four cubemaps, and generating a stereoscopic rendering visible from any viewing angle.

\section{Methodology}
\autoref{fig:Framework} illustrates the overview of the SCS projection pipeline. 

\paragraph{Rendering}
There are previous techniques for rendering a real-time monoscopic view on a cylindrical screen \cite{lorenz2009real}, but there are additional challenges to rendering accurate stereoscopic images for a cylindrical surface in real time. Many methods use pathtracing, which is computationally expensive for real time rendering \cite{ikkala2022tauray}. 

Performing a wrap around rendering for a cylindrical screen is more compute intensive because it requires performing traditional scene rendering multiple times. One method for curved surface rendering uses the technique stitching \cite{zhang2015casual} subdivides the screen into $N$ small segments along the curvature to approximate a planar slit. This results in $N$ render passes, where each pass draws a planar scene that is later being stitched together for the curvature image. For the case of no visible distortion, the stitching method would require one render pass per vertical pixel column, but approximations are often employed.

To create a stereoscopic view, we perform stereoscopic view rendering of the scene geometry. Our approach optimizes this process by rendering to four cubemaps, offset in the north, south, east and west by the interpupillary distance.

Before rendering each cubemap, bounds testing is done by checking which faces on the four cubemaps are visible from the user's head position, so that not all faces of every cubemap are rendered. When the player's head is close to the center, only a total of six faces across all cubemaps are rendered. This increases to a maximum of 20 potential faces for a head near the edge of the screen. Therefore, even at the upper bound, the required number of render passes remains substantially lower than that of the stitching-based approach.

\paragraph{Projection}

The four cubemaps are sampled to create a pair of stereo textures. The exact height and radius of the cylindrical screen is known, so the current head-tracked position is used to determine the viewing angle. To generate the stereo textures, the output display range is mapped linearly relative to the angle of the curved screen. For each fragment on the screen, the cubemap that is closest to the estimated position of the eye when looking in the direction of the fragment is sampled from.
For example, in the case of the northern portion of the projector, the northern face of the east and west cubemaps are used to create a stereo pair.

\section{Results and Discussion}

\paragraph{Implementation} We implemented our approach using Unity 6000.0.40f1 on a i7 17900K, 32GB RAM, NVIDIA RTX A5000 PC.

\paragraph{Performance} In our benchmark scene, the SCS renders at 60.1 FPS (16.6ms execution time), about 3x faster than a multi-view stitching based approach (22.8FPS, 43.8ms) \cite{zhang2015casual}. The slit-based camera approach approximates a flat plane by dividing the curved surface into slices of $270/32=8.4^o$.

The SCS projection pipeline is currently deployed on a cylindrical projection display ("The CAVERN") at Carnegie Mellon University Entertainment Technology Center. The CAVERN is a $270^o$ screen of a 3m radius and a height of 2.3m, and is used by students for their entertainment-design courses. The code base is available at  \url{https://bit.ly/4ceSWJF}, and video illustrations at \url{https://bit.ly/4cy0OWX}.

\paragraph{Future Work}

The SCS pipeline supports multiple users at the cost of accuracy by approximating rendering with a static viewing position at the center of the cylinder. For multiple head-tracked users, the view would need to be rendered once per user.

The current SCS implementation works on a single curve cylinder. For a multi-curved screen that approximates a sphere, the effects of pole merging need to be considered. A stitch-based solution \cite{zhang2015casual} will better resolve the artifacts of pole-merging, but the additional depth information on the cube map will allow us to recompose the cube as a 3DGS approximate, which allows a smoother mesh-to-mesh implementation \cite{lim2024projecting}.

SCS offers high performance and compatibility with existing render pipelines. This suggests that the technique is suitable for professional applications such as allowing designers to visualize decal designs on 3D models of cars using reverse projection \cite{lim2023reverse}.

\paragraph{Conclusions} 
Our SCS projection pipeline successfully renders and projects a world scene onto a cylindrical canvas with a stereoscopic view. Our pipeline provides an immersive viewing experience by optimizing the rendering for both eyes, and accounting for user position through off-center projection.

\begin{acks}
Thanks to Steve Audia who built the CAVERN, making this research possible, Faris Elrayes for the initial slit-based camera implementation, Ling Lin and Deyin Zhang for the art of the world scene, Yingjie Wang, Joshua Kaplan and Yun-Ni Tsai for their work on the CAVERN, and Drew Davidson for mentoring the CAVERN team.
\end{acks}

\bibliographystyle{ACM-Reference-Format}
\bibliography{references}


\end{document}